# An AI-enabled Agent-Based Model and Its Application in Measles Outbreak Simulation for New Zealand

Sijin Zhang, Alvaro Orsi and Lei Chen

Institute of Environment Science and Research, New Zealand

## Abstract

Agent Based Models (ABMs) have emerged as a powerful tool for investigating complex social interactions, particularly in the context of public health and infectious disease investigation. In an effort to enhance the conventional ABM, enabling automated model calibration and reducing the computational resources needed for scaling up the model, we have developed a tensorized and differentiable agent-based model by coupling Graph Neural Network (GNN) and Long Short-Term Memory (LSTM) network. The model was employed to investigate the 2019 measles outbreak occurred in New Zealand, demonstrating a promising ability to accurately simulate the outbreak dynamics, particularly during the peak period of repeated cases.

This paper shows that by leveraging the latest Artificial Intelligence (AI) technology and the capabilities of traditional ABMs, we gain deeper insights into the dynamics of infectious disease outbreaks. This, in turn, helps us make more informed decision when developing effective strategies that strike a balance between managing outbreaks and minimizing disruptions to everyday life.

## 1. Introduction

With the advancements in computational capabilities, the use of agent-based models (ABMs) has gained substantial traction across a diverse spectrum of scientific disciplines, particularly in those focused on understanding and analysing population dynamics such as public health, urban planning, agriculture, food safety and the related policy and governance (e.g., Bonabeau, 2002; Eberlen et al., 2017). ABMs delve into the micro-level interactions and behaviours of individual agents within a system, these agents can represent people, organisms, or entities, each distinguished by unique attributes, decision-making processes, and interactions with their environment and other agents. The collective impacts stemming from these agents can be thoroughly investigated, thereby providing valuable insights ranging from event-driven simulations to guiding principles for policy formulation.

Conventional ABMs, such as JUNE (Aylett-Bullock et al., 2021), typically adhere to an object-oriented design, which revolves around defining agents and specifying their actions (Cuesta-Lazaro et al. 2021, Chopra, Gel et al. 2021). In this design, each agent is characterized as an individual with distinct features, and interactions between agents occur in a sequential manner (Quera-Bofarull et al., 2023, Chopra et al. 2023). There is a highly evident and comprehensible logic when design the model in this way, however this means that each agent maintains separate memory allocations throughout the simulation, potentially leading to increased memory overhead and slower computation times (e.g., Chopra et al. 2023). Furthermore, calibrating the model often presents challenges and proves to be highly time-consuming, as a multitude of parameters can significantly influence the results (Kang and Choi 2010, Chang, Wilson et al. 2021).

To address the above issue, the concept of GradABM has been introduced by Ayush et al. (2022). This approach tensorizes the agents and their interactions within a Graph Neural Network (GNN, see Zhou et al., 2020 for more details) framework. The temporal progression of diseases is governed by a Long Short-Term Memory (LSTM) neural network (e.g., Hochreiter and Schmidhuber, 1997), which is characterized by a differentiable infection function, which takes into account the features and status of the agents such as susceptibility to the disease and the progression of the infection. The parameters of GradABM can be learned via the neural network backpropagation, in which the model adjusts its parameters in proportion to the error in its predictions. The GradABM model was established following the clinical setup for examining the administration of vaccination during the early outbreak of COVID-19 in London. This involved learning three key parameters: the overall infection rate, mortality rate and the initial infection percentage (AdityaLab, 2023). Extensive experiments demonstrate the efficacy of GradABM in rapidly simulating populations of millions by integrating deep neural networks and heterogeneous data sources.

The model we introduce in this paper, known as JUNE-NZ, is built by incorporating the concept of GradABM into the existing JUNE model (Aylett-Bullock et al., 2021). Compared to the original JUNE and GradABM, the model offers several significant advancements:

- Expanded Parameterization: A broader range of learnable parameters from neural networks.
- Measles Outbreak Simulation: We apply the model to simulate a measles outbreak in New Zealand and validate the model using real-world cases.
- Realistic Policy Environments for New Zealand: Our research extends to the establishment of more realistic policy environments tailored to the context of New Zealand.

In the subsequent sections of this paper, we provide detailed descriptions of our model setups in Section 2, followed by a thorough validation of the model's performance in Section 3. Section 4 delves into an analysis of potential policy measures aimed at mitigating measles transmission within school settings. Finally, Section 5 offers a concise summary of our research findings.

## 2. Model setups

### 2.1 Synthetic population

Following the guidance from the infectious diseases model JUNE (Aylett-Bullock et al., 2021), the synthetic population of New Zealand in JUNE-NZ contains two layers: population and interaction. The "population" data is derived from the 2018 New Zealand Census at the geography level of SA2, and it takes into account key demographic factors such as age, gender, occupation, ethnicity and various socio-economic indicators, aiming to create a comprehensive representation of the New Zealand population to the best extent possible. Additional information such as the immunisation coverage (obtained from New Zealand Ministry of Health) is also included in the "population" dataset. The "interaction" layer simulates social activities among agents, factoring in the frequency and intensity of contacts (denoted by $I_f$ and $I_i$, respectively) between individuals in various social settings like households, pubs, or cinemas. In this paper, the synthetic population is produced using the native JUNE model. Details can be found in Aylett-Bullock et al. (2021).

## 2.2 Methodology

### 2.2.1 Overview

The underlying disease model architecture is illustrated in the diagram below:

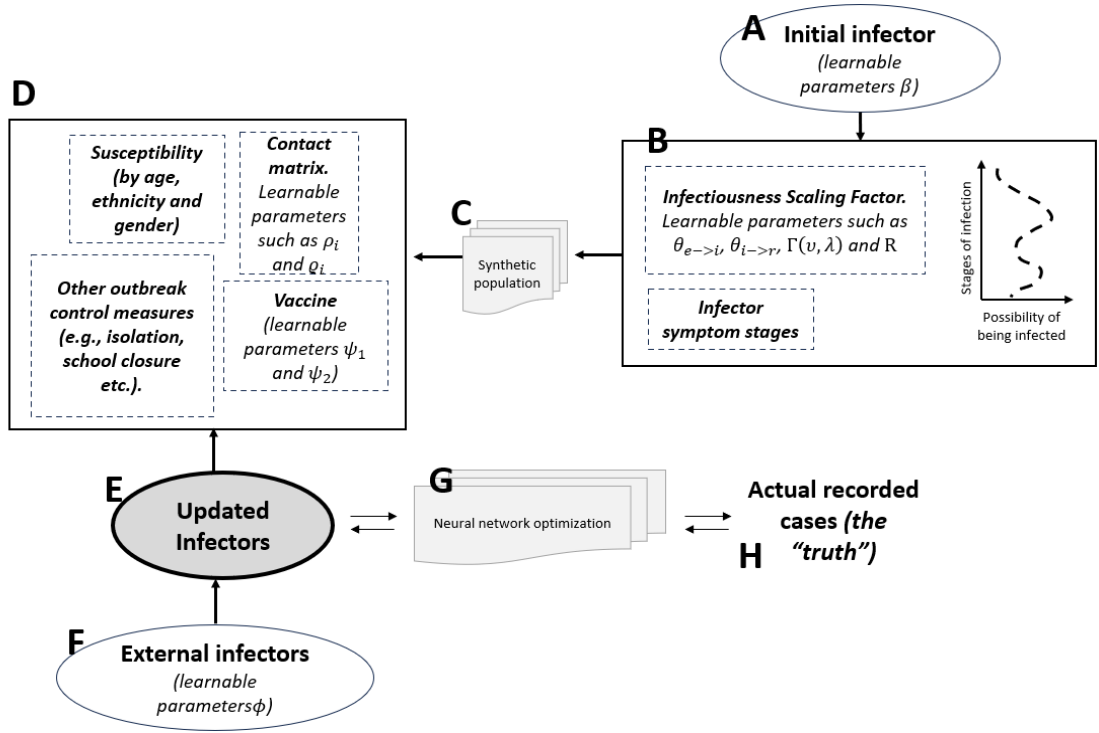

Figure 1: The model architecture of JUNE-NZ.

In our model, first we initiate a group of agents representing initial infectors (represented by the step A in the Figure 1). These agents' infectiousness is influenced by several factors such as the symptom stages (step B, see Section 2.2.2 for details). Depending on the infectiousness calculated by step B, the initial infectors disseminate the virus to other agents within various social settings. The likelihood of an agent contracting the infection is determined by several factors including age, ethnicity, gender, vaccine status and the implementation of outbreak control measures (step D and E, details can be found in Section 2.2.3 and 2.2.4). To account for uncertainties in the model, random infectors may be incorporated (step F, details in Section 2.2.5).

In the aforementioned steps, several learnable model parameters are involved. These parameters are iteratively adjusted during the neural network learning process (see Section 3.2 for more details) to minimize discrepancies (errors) between the total model infectors and the observed cases (steps G and H). The details of the learnable parameters in the model are listed in table 1:

Table 1: Key learnable parameters in JUNE-NZ

| Type | Description | |
|------|-------------|---|
| Infectors | The percentage of agents that get infected at the first-time step | $\beta$ |
| | The percentage of agents that get infected due to events outside the modelling area | $\phi$ |

| Infectiousness scaling factor | The number of days an agent takes to get infected after exposure of virus | $\theta_{e->i}$ |
| --- | --- | --- |
| | The number of days an agent takes to recover after being infected | $\theta_{i->r}$ |
| | The shape and scale parameters for the Gamma function, determining the infectiousness profile for an agent over time | $\Gamma(\upsilon, \lambda)$ |
| | The overall scaling factor for the infectiousness rate | R |
| Contact matrix | Contact intensity for different social venues. For example, for schools the parameter is represented as $\rho_{school}$. | $\rho_i$ |
| | Contact frequency for different social venues. For example, for schools the parameter is represented as $\varrho_{school}$. | $\varrho_i$ |
| Vaccine | The vaccine efficiency in reducing virus spread | $\psi_1$ |
| | The vaccine efficiency om reducing symptoms once a person is infected | $\psi_2$ |

Overall, the model is fine-tuned using real-world data by dynamically adjusting critical parameters related to infection timing, agents' contacts, and other relevant factors through a neural network framework. Once trained, this model becomes a powerful tool for exploring various policies aimed at curbing virus transmission within our society, and it enables us to conduct risk assessments for potential future outbreaks.

### 2.2.2 Initial infectors

As described in Section 2.2.1, a critical step in initializing the model is to designate a specific number of agents to be initially infected (step A in Figure 1). This process can follow a stochastic approach or adhere to user-defined criteria, such as confining the infected agents within a particular geographical region or originating from specific social venues. If the infectors are initiated via a stochastic process, to achieve differentiability in the initial infection process, we employ a *Gumbel SoftMax* function, denoted as $F_{G-S}$. Consequently, the count of individuals infected at the initial time step, denoted as $N_{t_0}$, can be expressed as follows:

$$N_{t_0} = F_{G-S}(\beta)$$

Here $\beta$ represents the proportion of agents subjected to infection at time $t_0$.

### 2.2.3 Infectiousness profile

The infectiousness profile is calculated when an agent is being infected (step B). This profile is represented by a trainable *Gamma* function $\Gamma(\upsilon, \lambda)$ that evolves according to the number of days elapsed after the agent's infection (an example is given in Figure 2). Here $\upsilon$ and $\lambda$ denote the shape and scale of the *Gamma* function, respectively, and they are updated via the model backpropagation.

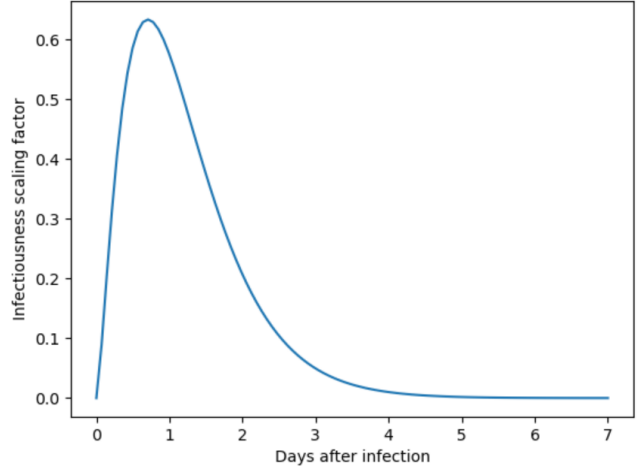

Figure 3: An example of the infectiousness scaling factor $\Gamma$ with shape $\upsilon$ and scale $\lambda$ parameters. In this example, $\upsilon = 2.41$ and $\lambda = 0.5$, while for JUNE-NZ, these parameters will be acquired and updated through the neural network backpropagation. The X-axis represents the number of days following infection and the Y-axis represents the infectiousness scaling factor.

Note that $\Gamma(\upsilon, \lambda)$ is constrained by another two learnable parameters $\theta_{e->i}$ (representing the number of days an agent takes to get infected after exposure of virus, where a susceptible agent would not get infected before $\theta_{e->i}$) and $\theta_{i->r}$ (representing the number of days an agent takes to recover after being infected, and infectiousness is set to zero for an infector after $\theta_{i->r}$).

### 2.2.4 Newly infected agents

The "newly infected agents" are produced by facilitating the exposure of susceptible individuals through their interactions with infected agents (step C, D and E in Figure 1). The susceptibility is dependent on multiple factors, including the settings of social venues (characterized by contact intensity and frequency denoted as $\rho_i$ and $\varrho_i$), the agents' vaccination status, outbreak control measures (see Section 2.2.6 for more details), and various machine learnable factors tied to the susceptible individuals' age, gender, ethnicity, and vaccination status.

The probability of agents being infected can be computed as follow:

$$P = \mathbf{R}\mathbf{H}_{attr}\frac{\rho_i}{\varrho_i}f(t, \theta_{e->i}, \theta_{i->r}, \Gamma(\upsilon, \lambda))$$

Where $\mathbf{R}$ represents the overall scaling factor for the infectiousness rate, which is learnable through model backpropagation. $\mathbf{H}_{attr}$ stands for the agent attributes matrix encompassing facets like age, sex, ethnicity, household, and the current symptom stage. $\rho_i$ and $\varrho_i$ are the contact intensity and frequency, respectively for different social venues. $f(t, \theta_{e->i}, \theta_{i->r}, \Gamma(\upsilon, \lambda))$ calculates the infectiousness considering timing parameters $t$ (the elapsed timestep since the agent is being infected), as described in Section 2.3.2.

### 2.2.5 Random infections

Stochastic infections can be introduced into the model over the entire simulation duration, excluding the initial time point $t_0$. This inclusion captures the inherent uncertainties within the model. The parameter $\phi$ signifies the rate at which agents are stochastically infected. Notably, $\phi$ is a time-dependant, trainable parameter that undergoes learning through model backpropagation. To ensure differentiability, the *Gumbel SoftMax* function is employed, and the count of agents impacted by this process can expressed as follows:

$$N_t = F_{G-S}(\phi)$$

### 2.2.6 Outbreak control

The outbreak control measures affect the susceptibility of an agent (described as part of $\mathbf{H}_{attr}$ in Section 2.3.3). Additionally, one of the main targets for this model is to investigate different policies for reducing the virus transmission in a community. For example, for measles, the Ministry of Health (MoH) New Zealand has developed a comprehensive set of case management policies, encompassing guidelines for notification procedures, case and contact management, and vaccination recommendations. To ensure the model remains relevant to real-world settings and enables the evaluation of potential impacts under various scenarios, the following policies are incorporated in JUNE-NZ:

Table 2: Outbreak control measures implemented in JUNE-NZ.

| Full name | Abbreviation | Description |
|---|---|---|
| Isolation of Confirmed Case | ICC | Confirmed measles cases in the model must self-isolate for at least 4 days post-rash. |
| Quarantine of Exposed Contacts | QEC | In the model, unvaccinated individuals exposed to measles must quarantine for 7-14 days post-exposure. Vaccinated individuals are exempt. |
| School Closure | SC | School closures (1-2 weeks) may occur in the model if a measles case is identified. |
| Vaccination Campaign | VC | In the model, a vaccination campaign may be initiated in the event of a nearby measles case detection. |

It is noteworthy that certain parameters influence the implementation of outbreak control measures. Notable examples include the compliance rate of isolation, which represents the percentage of individuals adhering to self-isolation requirements, and the contact tracing rate, which reflects the effectiveness in tracking individuals exposed to infectors. These parameters can be user-specified to explore the impact of different policies within the constraints of real-world scenarios.

Building upon the above discussion, the total number of infected agents at step E in Figure 1 can be determined through the following calculation:

$$N_t = N_{t_0} + [N_{t-1} + (K_{sus,t-1}C_{ctl}P_{t-1})]$$

Here $K_{sus}$ denotes the total number of susceptible agents at time step $t-1$. $C_{ctl}$ represents the influence of implemented outbreak controls, and $P_{t-1}$ represents the probability of a susceptible agents becoming infected.

# 3. Model simulation and validations

## 3.1 Experiment design

In 2019, New Zealand experienced a significant resurgence in measles cases, marking the most substantial outbreak in over two decades. Between January and late September 2019, the country recorded more than 1,500 confirmed cases of measles, with over a third of this case necessitating hospitalization. These cases were distributed across 16 out of the 20 District Health Boards (DHBs) in New Zealand. Most of these cases were effectively managed through robust public health interventions. However, a notable exception emerged in the Auckland area, particularly centred around the County Manukau District DHB region.

This focuses on modelling the outbreak's peak between week 26 (starts from June 24, 2019) and week 51 (ends on December 20, 2019) in Manukau DHB. As delineated in Section 2, the JUNE-NZ model encompasses a multitude of parameters, a subset of which is self-learnable through the mechanism of neural network-guided backpropagation.

## 3.2 Model training

The primary objective of this section is to enable the calibration of intrinsically adjustable parameters. This process is designed to achieve alignment between the simulated case numbers and the actual case numbers reported or observed within the Manukau District Health Board (DHB) region. The progression of loss across 100 iterations, obtained through the use of the Stochastic Gradient Descent (SGD) loss function (Ruder, 2016) in the optimization process, is shown in Figure 5.

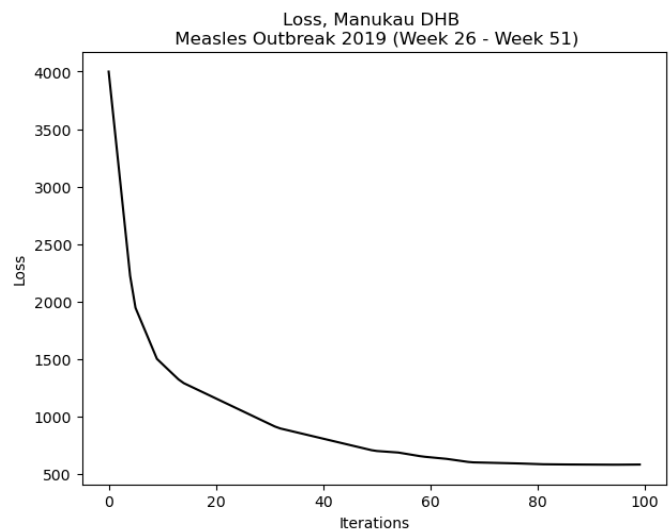

Figure 4: The loss of JUNE-NZ for simulating the measles outbreak in Manukau DHB occurred between week 26 and week 51.

Referencing Section 2, the model optimization involves learning and updating multiple parameters. These parameters are initialized using the *Xavier Uniform Initialization* Technology to facilitate effective weight initialization. Subsequently, the parameters are coherently updated to minimize the loss between the estimated and observed cases. An example is showcased in Figure 5, showing the evolving trend of the initial infection rate, denoted as $\beta$, across iterations. Commencing around 0.0015%, the value of $\beta$ stabilizes at approximately 0.003% after roughly 40 iterations. Considering that with the population of Manukau DHB being around 530,000, our simulation suggests that roughly 16 residents are already being infected before the model's initiation.

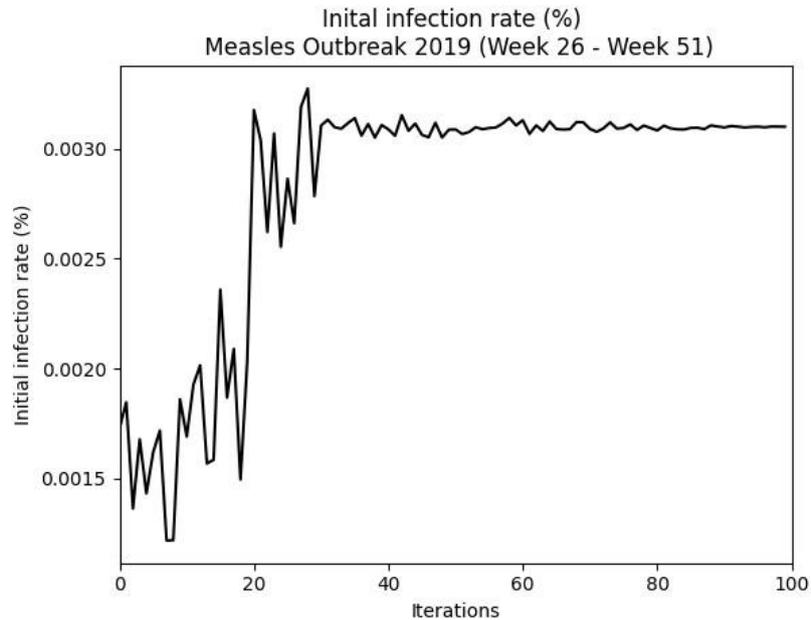

Figure 5: The evolution of the initial infection rate (%) throughout the model optimization process.

The parameters also exhibit temporal dependency, with the temporal information being extracted from the hidden states of an integrated LSTM neural network within JUNE-NZ. These temporal updates are essential as they capture the natural progression of the disease and the evolving societal responses over the simulation period. For example, Figure 6 shows how the virus's infectiousness evolves over the simulation timeframe. The initial scaling factor is normalized at 100%. It becomes evident that infectiousness gradually reduces, exhibiting relative stability from week 29 to week 43 of 2019, followed by a subsequent drop to 97% towards the simulation's conclusion.

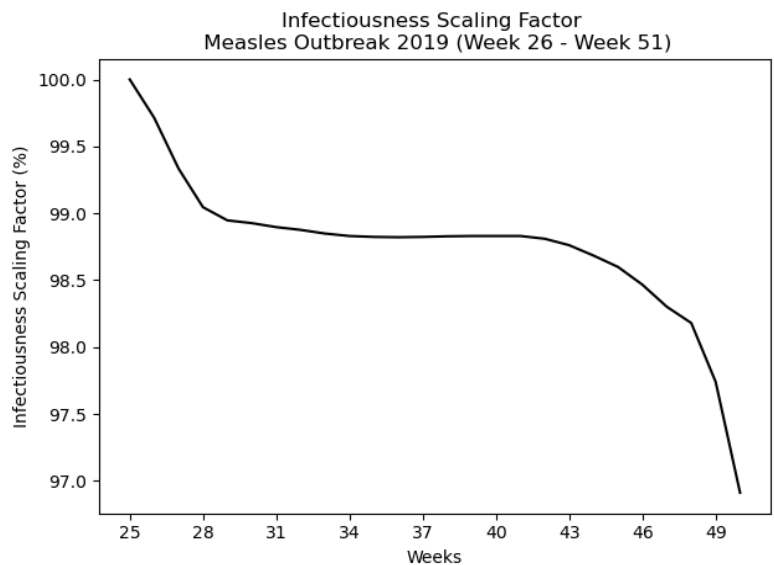

Figure 6: Infectiousness scaling factors for the measles outbreak between week 26 and week 51 of 2019 in the Manukau DHB.

Overall, the primary objective of model training, particularly within the scope of JUNE-NZ, is to fine-tune the disparity between the simulated case numbers and the actual observed data. To achieve this, we utilize SGD as a means to optimize our cost function. This optimization process entails a continuous update of parameters to minimize the variance between the projected and observed cases. This is accomplished within the framework of a GNN. To account for temporal dependencies, we incorporate a Long Short-Term Memory (LSTM) neural network into our model. The culmination of this process unveils the intricate dynamics of virus transmission.

### 3.3 Simulation validation

Figure 7 illustrates the simulated incidence of measles cases within the Manukau DHB alongside the corresponding reported cases documented in EpiSurv, the New Zealand National Communicable Diseases Database administered by the Institute of Environmental Science and Research (ESR). To account for uncertainties inherent in the model and input data, a total of 100 ensemble runs were executed, each with distinct initial infected agents and perturbed agents' interaction behaviours.

The simulation forecasts a cumulative total of 1167 measles cases for the period of JUNE-NZ, this figure being an average across all ensemble members. This contrasts with the 1024 cases reported by the DHB during the identical timeframe. A noteworthy observation from the simulation is the identification of the outbreak's apex in week 36, where the simulated case count escalates to 119, compared to the actual reported case count of 91. Following this simulated peak, the model predicts a decrease in cases, while it lags behind the rate observed in the reported cases.

Dung the peak weeks of the case count, the model simulation exhibits substantial uncertainties, which emphasizes the model's sensitivity to social dynamics within this period in the Manukau DHB region. It is important to highlight that the model has incorporated standard outbreak control measures as detailed in Section 2.2.6. For example, we have set the compliance rate of ICC within a range of 60% to 80%, and this range significantly contributes to the spread of the number of cases simulated in the model when the case number is high. It's crucial to note that these measure ranges are manually determined and adjusted, given the absence of concrete evidence or statistical data.

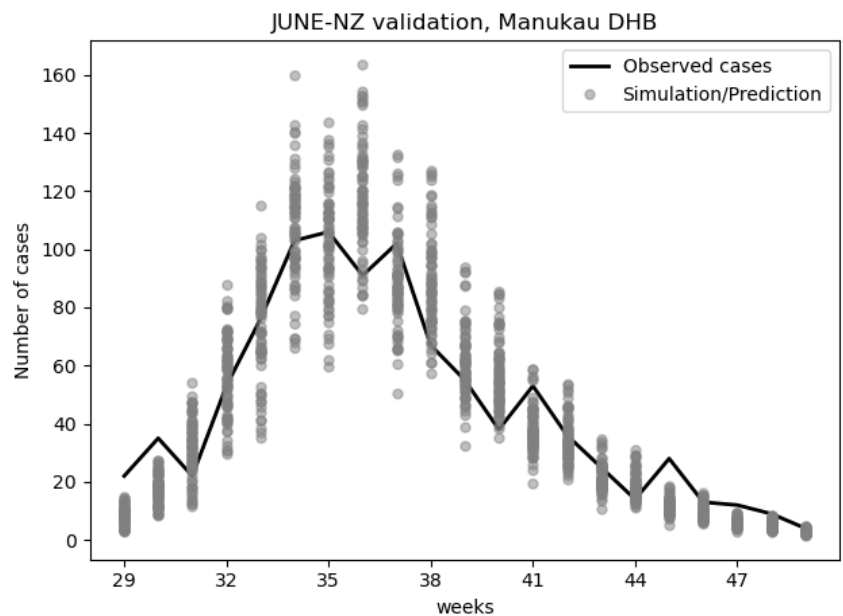

Figure 7: The simulated and reported number of cases in Manukau DHB between week 26 and week 51. The back solid line indicates the reported measles cases in the ESR notifiable disease database.

Figure 8 presents a comparative analysis of the simulated and observed cases, focusing on the distributions of ethnicity (Figure 8: left) and age (Figure 8: right). The actual data, represented by a black cross, is calculated as a percentage based on the recorded cases. Each ensemble member's result from the model is denoted by a grey dot.

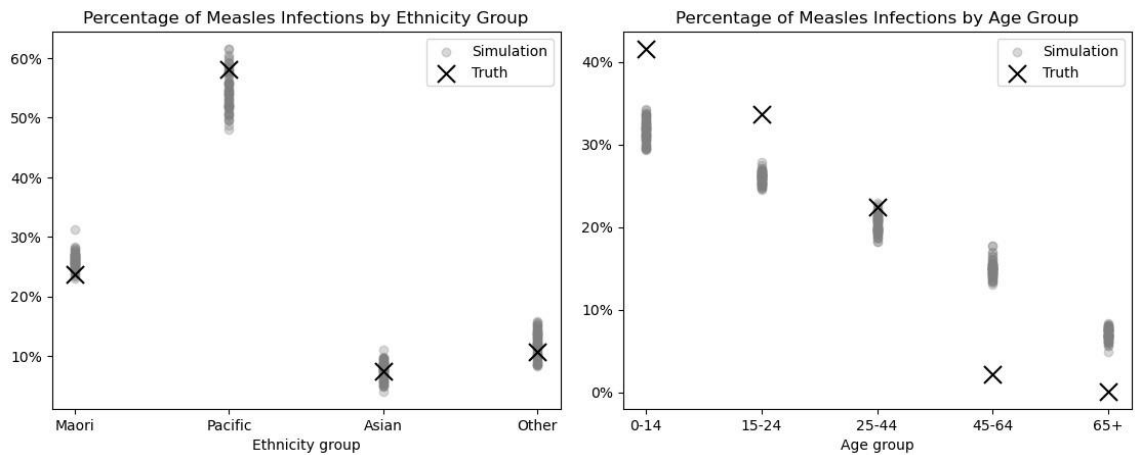

Figure 8: Percentage of measles infections by ethnicity group (left) and age group (right)

In the left panel of Figure 8, the model demonstrates a high degree of accuracy in simulating the infection rates among the Māori and "Other" ethnic groups, which includes Europeans and MELAA. However, it exhibits a notable underrepresentation of Pacific infectors, with a discrepancy exceeding 15%. In contrast, the model significantly overestimates the infection rates among the Asian population. From Figure 8 (right), it becomes evident that the model tends to underestimate the infection rates among younger individuals, specifically those under the age of 24. Conversely, it overestimates the infection rates for older individuals, particularly those aged 45 and above. This dichotomy suggests potential areas for refinement in the model's demographic sensitivity. The potential sources of the above discrepancies within the model, and the prospective ways to improve it are comprehensively discussed in Section 4.

The model under discussion has been trained using target data sourced from the Manukau DHB. This makes it particularly suited for simulating a variety of policy scenarios and demographic changes. It can provide insights into how these changes might influence virus transmission within the community, should an event akin to the 2019 outbreak occur in the future. The value of such modelling lies in its potential to serve as a tool for policy investigation, thereby informing the development of effective strategies for disease control.

Additionally, the proposed model holds potential for predicting future outbreaks. Conventionally, such predictions are facilitated by a variety of predictive models, with regression models being a common choice. However, the task becomes considerably challenging in the context of measles due to the scarcity of sufficient data. An Agent-Based Model (ABM) could potentially address this issue, owing to its inherent ability to incorporate well-defined social networks that exhibit variability across different locations. A model that has been trained in one specific location, with parameters that accurately describe the dynamics of the virus (for example, the infectiousness of the virus contingent on the symptom stage of the infector), could provide valuable insights into the transmission patterns of the virus in a different location by assuming the consistent virus characteristics.

In order to assess the model's effectiveness in replicating real-world outbreak scenarios across different regions, we utilized the model trained from Manukau DHB to predict/simulate the scenarios of the outbreak that occurred in Auckland DHB and Capital and Coast DHB (CC-DHB). Note that to account for regional variations in the initial number of infectors at the model's start time, and in light of the absence of statistical data, the seed infectors for both the DHBs are manually adjusted to optimize model performance.

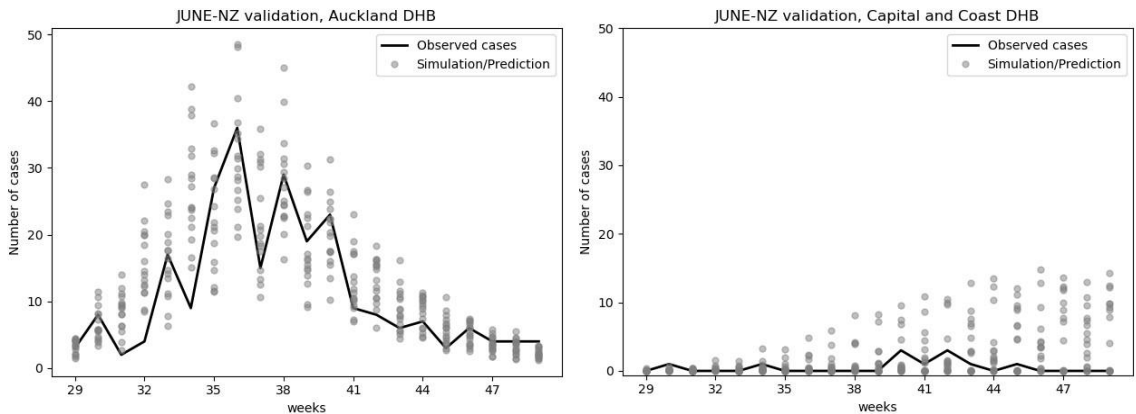

Figure 9: The simulation for the number of measles cases in Auckland DHB (left) and Capital and Coast DHB (right) using the model parameters trained in the Manukau DHB.

The model demonstrated a good performance during the simulation period for the Auckland DHB, accurately capturing the apex of the outbreak. However, when applied to the Manukau, the ensemble members from the Auckland DHB exhibited substantial uncertainties, particularly during periods of high observed case numbers. In contrast, the model's performance for the CC-DHB during the same period was suboptimal. Although the model was capable of giving a smaller number of infected cases for the CC-DHB compared to Manukau or Auckland, it appeared to overestimate the outbreak within the target region of interest, indicating a prolonged outbreak peak period with weekly cases ranging between 15 and 18. In reality, the highest number of observed cases only occurred during weeks 40 and 42, with the case number being less than 10.

## 4. Discussion

In this section, we highlight some constraints inherent to the current model and propose potential avenues for future development:

- First, the model presented in this paper demonstrated proficiency in rapidly replicating past outbreak events, thereby facilitating the study of different policy impacts and societal risk alterations, however its predictive capabilities were found to be limited (see Section 3.3 for more details). This limitation was particularly evident when the target area exhibited a demographic distribution drastically different from that of the training dataset. The scarcity of historical data, especially for diseases like measles in New Zealand, further compounds this issue.

- Second, our experiments revealed that the initial seed infectors utilized in the model significantly influence the accuracy of the modelling. While this parameter can generally be learned from the neural network backpropagation given a reasonable range, challenges arise when applying the trained model to a new region or a new simulation period. To enhance the capability of modelling/predicting of virus transmission, it is imperative to augment our case tracking capabilities and utilize this as the initial condition for the model.

- Third, the model does not support cross-region modelling, such as virus transmission between Auckland and Waikato, due to the absence of pertinent travel data. This constraint hampers the model's capacity to provide advice and assess policies at a nationwide level. The authors are currently liaising with the New Zealand Ministry of Transport to investigate the feasibility of accessing the relevant dataset and assessing the quality of specified dataset.

- Fourth, this study underscores the importance of ensemble-based simulation over deterministic simulation to address model uncertainties. However, the current model generates ensembles through embedded differentiable stochastic processes, such as the use of *Gumbel Softmax* function. As a future direction, it is essential to incorporate the capability of perturbing the synthetic population, particularly the interactions among agents, to mirror the complex dynamics within our community.

- Finally, one of the most critical aspects of enhancing the model is the collection of more granular information for the synthetic population. For instance, the current model categorizes ethnicity broadly into Māori, European, Asian, MELAA, and others. However, significant social dynamics may exist within these broad ethnic groups. To facilitate more targeted policy simulation, it is recommended to disaggregate these ethnic groups into more detailed categories.

## 5. Conclusion

This study presents an innovative agent-based model (ABM), JUNE-NZ, that leverages the capabilities of Graph Neural Networks (GNN) and Long Short-Term Memory (LSTM) networks. This AI-enhanced ABM offers a significant improvement over traditional models by enabling rapid and potentially real-time parameter calibration, particularly useful during ongoing outbreaks, and efficient scalability for large populations.

The model's parameters are optimized using the Stochastic Gradient Descent (SGD) loss function through neural network backpropagation. These parameters were trained and calibrated using real-world infectious disease data, specifically focusing on measles. The model was then applied to simulate the 2019 measles outbreak in the Manukau District Health Board (DHB), producing encouraging results.

The same parameters derived from the Manukau DHB were utilized to run simulations for the Auckland and Capital and Coast DHBs. While the Auckland DHB simulation yielded reasonable outputs, the results from the Capital and Coast DHB were generally unsatisfactory. The paper also discusses several potential enhancements to improve the model's performance.

In conclusion, the model introduced in this paper represents a promising tool for rapid response modelling and policy simulation in the event of a measles outbreak, thereby providing evidence-based recommendations for planning strategies. Furthermore, the model's potential extends to its application in other communicable diseases, marking the next step in its development.